\documentclass[journal=jacsat,manuscript=article]{achemso}

\usepackage[version=3]{mhchem} 
\usepackage{natbib}

\usepackage{amssymb}
\usepackage{array}
\usepackage{makecell} 
\usepackage{xcolor}
\usepackage[symbol]{footmisc}
\usepackage[utf8]{inputenc} 

\newcolumntype{P}[1]{>{\centering\arraybackslash}p{#1}}

\author{Hongyu Gao}
\affiliation[Saarland University]{Department of Materials Science \& Engineering, Saarland University, Campus C6.3, 66123 Saarbr\"ucken, Germany}
\email{hongyu.gao@uni-saarland.de}
\phone{+49 681-302-57458}
\author{Sergey Sukhomlinov}
\affiliation[Saarland University]{Department of Materials Science \& Engineering, Saarland University, Campus C6.3, 66123 Saarbr\"ucken, Germany}

\title[]{Is Ultralow Friction on Graphite Sustainable in Contaminated Environments?}

\abbreviations{IR,NMR,UV}

\begin{document}








\begin{abstract}

Structural lubricity arises typically at incommensurate, well-defined dry contacts where short-range elastic instability is significantly mitigated.
However, under ambient conditions, airborne molecules adsorb onto solid surfaces, forming an intervening viscous medium that alters interfacial properties.
Using molecular dynamics simulations with a newly parameterized interfacial potential, we investigate the preservation of ultralow friction on graphite with physisorbed $n$-hexadecane (HEX) as a model contaminant. 
Our findings reveal that a well-ordered monolayer of HEX molecules strongly adheres to the graphite surface, replicating its lattice structure and maintaining solid-like behavior, which leads to orientation-dependent shear stresses---an effect absent on a gold (111) surface.
As the contaminant film\footnote[1]{A film can consist of one or more molecular layers} thickens, this orientation effect diminishes.
Additionally, as coverage increases from zero to one monolayer, the shear stress-velocity relationship transitions from Coulomb to quasi-Stokesian and then to quasi-Coulomb, highlighting the role of molecular displacement in high-velocity dissipation.
Despite a hundredfold increase in shear stress compared to dry sliding, superlubricity on graphite persists under ambient conditions, enhancing our conventional understanding of structural lubricity.

\end{abstract}


\section{Introduction}

Structural lubricity~\cite{Mueser2004EPL} refers to an ultra-low friction state that arises from the systematic annihilation of interfacial lateral forces due to lattice mismatch.
In this context, interfacial atoms can be treated as the smallest contact units, with their energy states, driven by thermal motion, distributed stochastically across a potential energy landscape.
Stokesian damping~\cite{Adelman1976JCP}, which describes a linear increase in friction with sliding velocity, results from small perturbations of these interfacial atoms.
This leads to primary energy dissipation as the atoms vibrate around their equilibrium positions within the lattice, without engaging in long-range elastic or quasi-discontinuous motion under shear forces.
A scaling argument, based on the law of large numbers, dictates a sub-linear relationship between friction and contact area~\cite{Mueser2001PRL, Dietzel2013PRL}: $F\propto A^n$, where the exponent $n$ is capped at 0.5 and varies depending on the degree of lattice registry and geometric nature of the contact, including contact lines~\cite{Mueser2004EPL, de-Wijn2012PRB, Gao2022FC}.
While this low-friction state is highly desirable, it can become unstable when the elastic correlation length is exceeded~\cite{Sharp2016PRB, Monti2020ACSN}.
Beyond this threshold, structural defects, such as dislocations, can cause the sliding surfaces to interlock, significantly increasing shear stresses up to the Peierls stress limit~\cite{Sharp2016PRB}.

Under ambient conditions, airborne molecules such as water and short alkanes tend to adsorb onto solid surfaces, forming an orderly arranged monolayer that aligns with the underlying solid lattice~\cite{He1999S}.
With prolonged exposure, this monolayer can develop into a nanometer-thick multilayer film~\cite{Annett2016N}, as indicated by an oscillatory density profile resulting from the liquid’s wavenumber-dependent compressibility~\cite{Gao2020JCIS}.
The anisotropic behavior of confined liquids, characterized by high in-plane molecular ordering, enables a load-bearing ability to a certain degree while still allowing molecular diffusion~\cite{Gao2020JCIS, Gao2022L, Huang2023NC}.
Interestingly, the presence of such an adsorption layer does not necessarily impede superlubricity~\cite{Cihan2016NC, Deng2018N, Oo2024}, with friction only mildly increasing alongside observations of rejuvenation, aging, and friction switches~\cite{Oo2024}.
This suggests that earlier nano-manipulation experiments may have overlooked the existence of these layers, as contamination is unlikely to be fully eliminated, even under nominal ultra-high vacuum conditions.
Unlike the constant velocity gradient typical of Couette flow, where molecules must overcome energy barriers due to steric hindrances, the shear plane in the boundary lubrication regime is located at heterojunctions with the lowest shear strength, exhibiting solid-like behavior.
Consequently, when sliding occurs over such a viscoelastic medium, the frictional dependence on sliding speed is expected to be sub-linear~\cite{Mueser2020L}.

Building on our previous studies~\cite{Gao2022FC, Gao2024TL}, we focus on highly oriented pyrolytic graphite (HOPG) due to its superior performance as a solid lubricant.
The aim is to examine the extent to which superlubricity is maintained in the presence of linear alkanes, specifically $n$-hexadecane (HEX), as a representative adsorption contaminant.
To replicate the experimentally observed morphology of parallel stripes of HEX, with their longitudinal axes aligned parallel to the zigzag direction of HOPG, we reparameterized the interatomic interaction between HOPG and HEX based on density-functional theory (DFT) calculations.
To validate this new parameter set, we calculated binding and desorption energies and compared them with experimental measurements.
We then investigate how topography influences molecular in-plane ordering during liquid confinement.
Finally, we study energy dissipation both in area-filling buried interfaces and at contact lines within the boundary lubrication regime.

\section{Methodology}

Molecular desorption, liquid confinement, and boundary shearing are investigated using molecular dynamics (MD) simulations, employing model systems that incorporate Au(111) and graphite (0001) (Gr) as solid surfaces, with $n$-hexadecane (HEX) as the adsorbent.
To ensure clarity, detailed geometric information and operational conditions for each model system are provided in the relevant sections.
The interactions of Au, Gr, and HEX themselves are described using the EAM~\cite{Zhou2004PRB}, AIREBO~\cite{Stuart2000JCP} potentials, and the L-OPLS~\cite{Price2001JCC, Siu2012JCTC} force-field, respectively.
For cross-interactions, Morse~\cite{de-la-Rosa-Abad2016RSCA} and Lennard-Jones~\cite{Pu2007N} potentials are utilized to describe Au-Gr and Au-HEX interactions, respectively.
The interaction between Gr and HEX is reparameterized, with further details provided in the Results section.
Unless stated otherwise, the system temperature is maintained at 300 K using a Langevin thermostat with a damping factor of 0.1 ps.
The thermostat is applied in all directions except during sliding simulations, where it is restricted to the $y$-axis, perpendicular to the sliding direction.
The simulation timestep is set to 1 fs for all cases.
All MD simulations are carried out using the open-source code LAMMPS\cite{Thompson2022CPC}.
To determine the optimal MD data output frequency, a time autocorrelation function (ACF) analysis is performed, with mean values and standard errors calculated based on uncorrelated data sets extracted.

\section{Results and discussion}
\subsection{Interfacial Potential Development}

The interfacial interaction between graphite (Gr) and $n$-hexadecane (HEX) is described using the Buckingham potential~\cite{Buckingham1938PRSL}, formulated as:
\begin{equation}
E_{\rm int} = A{\rm e}^{-r/\rho}-\frac{C}{r^6},
\label{eqn:buck}
\end{equation}
where $A$, $\rho$, and $C$ are fitting parameters optimized through force matching (FM).
This potential provides a more precise representation of short-range repulsive interactions with an exponential form, derived from the interpenetration of closed electron shells, making it more suitable for modeling boundary shearing~\cite{Kong2009PCCP, Mueser2022MS}.
Reparameterizing these potential parameters, rather than relying on mixing rules~\cite{Lorentz1881AP, Berthelot1898CR} for the non-bonded Lennard-Jones (LJ) parameters from the OPLS force field, is crucial because the latter, designed for bulk phase properties, fails to accurately capture the interfacial interactions, leading to a significant underestimation of the energy corrugation barrier.
Two sets of parameters for ${\rm C_{Gr}}-{\rm C_{HEX}}$ and ${\rm C_{Gr}}-{\rm H_{HEX}}$ were determined by reproducing DFT-calculated static molecular forces ($f$) and binding energy ($\Delta E$, defined in Eq.~\ref{eqn:binding_energy}) using a single-molecule-on-graphene model system.
Sixteen configurations ($N_{\rm conf}=16$) were considered, encompassing a range of both in-plane and out-of-plane relative positions between the two substances.

First-principles DFT calculations were performed using the Gaussian and plane-waves methodology~\cite{VandeVondele2005CPC, Gerald1997MP} within the {\it Quickstep} module of the open-source software CP2K~\cite{Hutter2014CMS, Kuhne2020JCP}.
A plane-wave cutoff of 900 Hartree was applied to achieve a force convergence tolerance of 10$^{-5}$ atomic units (a.u.).
$\kappa$-point sampling was employed along with periodic boundary conditions in all three dimensions.
To minimize empirical influences, a van der Waals density functional (vdW-DF)~\cite{Dion2004PRL} was used, combining exchange at the generalized-gradient approximation (GGA) level, correlation at the local density approximation (LDA) level, and a non-local correction for van der Waals interactions.
Carbon and hydrogen atoms were described using a triple-$\zeta$ Gaussian basis set~\cite{VandeVondele2007JCP}, paired with Goedecker-Teter-Hutter (GTH) pseudopotentials~\cite{Goedecker1996PRB, Hartwigsen1998PRB}.
The results closely align with those obtained using other widely-used methods, such as Perdew-Burke-Ernzerhof (PBE)~\cite{Perdew1996PRL} with Grimme's DFT-D3~\cite{Grimme2010JCP} corrections for dispersion interactions.

In FM, the discrepancy between DFT and force-field (FF) calculations was quantified using a penalty function $\chi^2$, defined as:
\begin{equation}
\chi^2 = \frac{1}{D} \frac{\sum_{\alpha =1}^D \omega_{\alpha} \chi_{\alpha}^2}{\sum_{\alpha =1}^D \omega_{\alpha}},
\end{equation}
where $D~(D=4)$ represents the dimensionality including three force components plus energy, and $\omega$ (with $ \omega_{f_x} = \omega_{f_y} = 5\times\omega_{f_z} = 0.1\times\omega_{\Delta E}=1.0$) denotes the weighing factors determined based on their respective value spans.
The term $\chi_{\alpha}^2$ is defined as:
\begin{equation}
\chi_{\alpha}^2 = \frac{1}{N_{\rm conf}} \sum_{i=1}^{N_{\rm conf}} \frac{(\psi_{i,\alpha}^{\rm FF}-\psi_{i,\alpha}^{\rm DFT})^2}{(|\psi_{i,\alpha}^{\rm DFT}|+\Delta)^2},
\end{equation} 
where $\Delta=\underset{i,\alpha}{\rm max} (|\psi_{i,\alpha}^{\rm DFT}|)$ serves as a denominator adjustment to mitigate the impact of small values on fitting stability, and $\psi=\{f,\Delta E$\} represents the physical properties being compared.
The minimization of $\chi^2$ was executed using Monte Carlo-based simulated annealing~\cite{Kirkpatrick1983S}. 
The fitting parameters were accepted upon meeting the convergence criterion of $\chi^2 \leq 10^{-3}$, where $\chi^2=0$ indicates perfect agreement between FF and DFT calculations.

\begin{figure}[ht]
\centering
\includegraphics[width=0.7\textwidth]{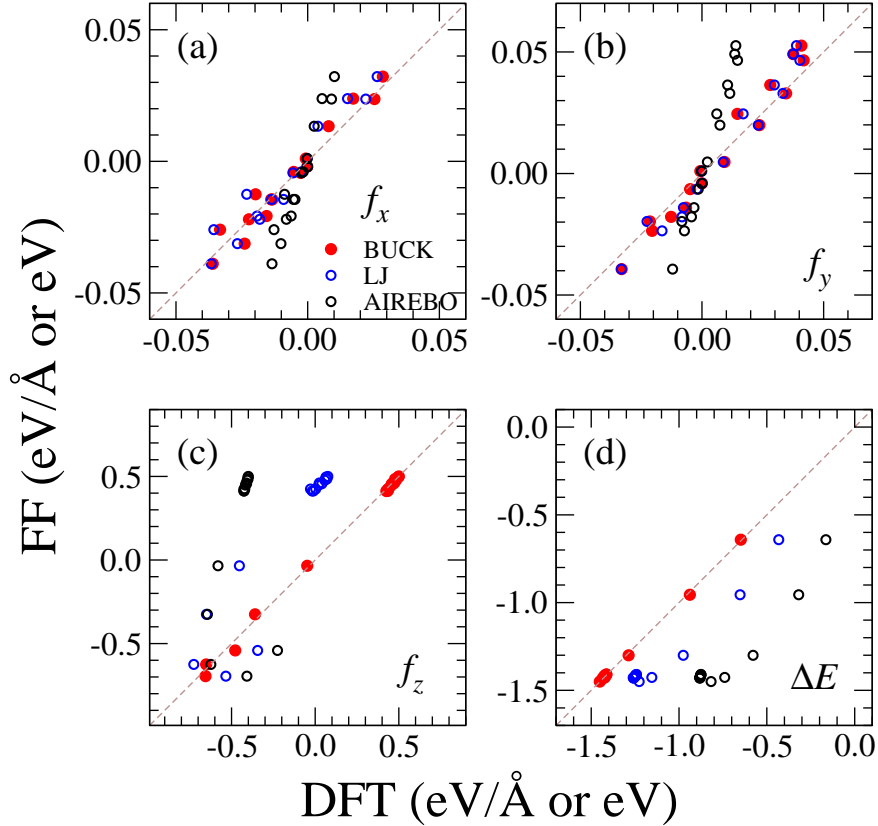}
\caption{
Comparison between DFT- and FF-calculated molecular forces ($f$) along the a) $x-$, b) $y-$, and c) $z-$direction, and d) binding energy ($\Delta E$). 
When $[\psi=\{f,\Delta E\}]^{\rm FF} \equiv \psi^{\rm DFT}$, the data points align precisely with the dashed lines. 
The LJ parameters were obtained using mixing rules on the all-atom L-OPLS parameters.
}
\label{fig:force_fitting}
\end{figure}

Figure~\ref{fig:force_fitting} compares the results from DFT and FF calculations, showing that the newly parameterized Buckingham potential (red solid circles) outperforms others in accurately reproducing both molecular forces and energies.
The corresponding potential parameters are listed in Table~\ref{tab:buck_parameter}.
Notably, the out-of-plane forces ($f_z$) estimated with the mixed OPLS-LJ potential were significantly underestimated (blue open circles in Fig.~\ref{fig:force_fitting}c), especially in the repulsive regime.
As shown in Fig.~\ref{fig:force_fitting}d and supported by our previous study~\cite{Oo2023TL}, the normal force component influences the energy barrier, thereby affecting lateral motion.
%
Predictions based on the AIREBO potential (black open circles) are notably less accurate in this context.

\begin{table}[ht]
\centering
\begin{tabular}{P{2.5cm}P{2.5cm}P{2.5cm}P{2.5cm}}
\Xhline{1.0pt}
$i-j$ & $A_{ij}$ (eV) & $\rho_{ij}$ (\AA) & $C_{ij}$ (eV$\cdot$\AA$^6$) \\
\hline
$\rm C_{Gr}-C_{HEX}$ & 1.75 & 0.09 & 57.13 \\
$\rm C_{Gr}-H_{HEX}$ & 132.66 & 0.34 & 0.00 \\
\Xhline{1.0pt}
\end{tabular}
\caption{Optimized Buckingham parameters for Gr-HEX interactions}
\label{tab:buck_parameter}
\end{table}

Reproducing collective properties like energy differences and molecular forces is relatively straightforward compared to reproducing per-atom forces, which require much complex functional forms that we were unable to identify.
Specifically, we recognize our limitations in developing bond-order many-body potentials, including parameterizing the embedded-atom method (EAM)~\cite{Daw1984PRB} and Steele~\cite{Steele1973SS} potentials, to account for possible many-body effects resulting from CH/$\pi$ interactions~\cite{Tsuzuki2008PCCP} that induce electron density changes upon approach.
At this stage, we disregard the electrostatic interaction between Gr and HEX, assuming the graphene sheet is electronically neutral.
We did not pursue machine-learning techniques, despite their promise, as the results might not align with established physical principles.
Although imperfections exist, our approach achieves the initial objective of developing simple yet accurate means for modeling interfacial dynamics on a large scale with minimal computational cost.

\subsection{Physisorption of $n$-hexadecane on Graphene}

To validate the fitting parameters, we compared the binding energies ($\Delta E$) required to dissociate a single HEX molecule from a Gr substrate with experimental data~\cite{Paserba2001PRL, Gellman2002JPCB}.
These energies were determined based on two equilibrium energy states, formulated as:
\begin{equation}
\Delta E = \langle E_{\rm bound} \rangle - \langle E_{\rm sub} + E_{\rm HEX}^{\rm des} \rangle
\label{eqn:binding_energy}
\end{equation}
Here, $E_{\rm bound}$ represents the potential energy of the system when the target molecule is adsorbed on Gr, while $E_{\rm sub}$ and $E_{\rm HEX}^{\rm des}$ correspond to the potential energies of the substrate (including any remaining HEX molecules) and the desorbed molecule in isolation, respectively.
The notation $\langle \rangle$ denotes a time average taken over a minimum duration of 10 ns.
The coverage density ($\it\Gamma$) is determined by calculating the total in-plane occupation of HEX molecules, denoted as $mA_{\rm HEX}^{\rm single}$, where the projected area of a single HEX molecule ($A_{\rm HEX}^{\rm single}$) is approximately 105 \AA$^2$~\cite{Londero2012JPCM}.
Here, $m$ represents the total number of molecules, which ranges from 1 to 30, corresponding to a $\it\Gamma$ varying from 0.03 to 0.77 monolayers (ML).

\begin{figure}[ht]
\centering
\includegraphics[width=0.55\textwidth]{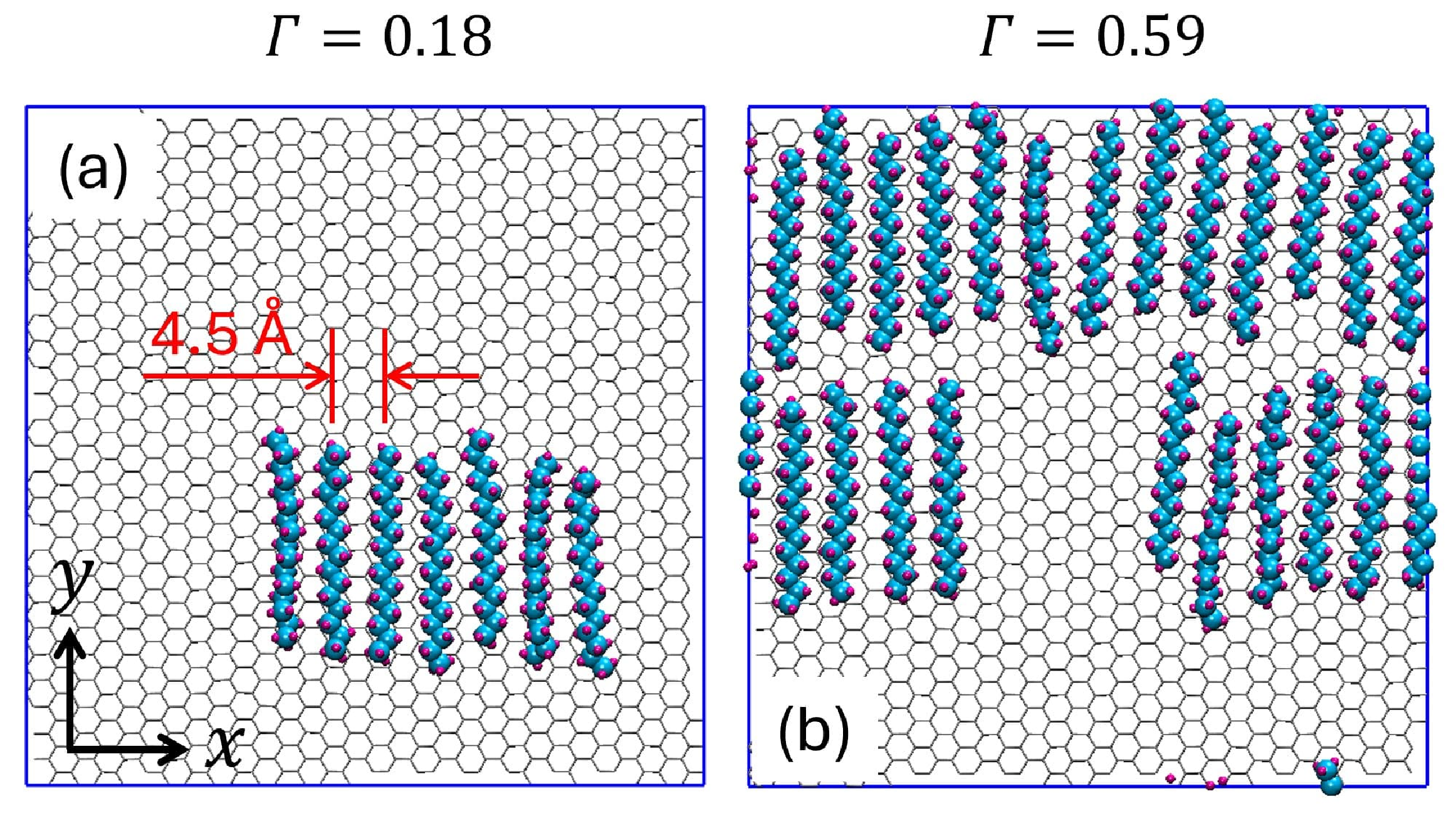}
\includegraphics[width=0.38\textwidth]{pr_orient.eps}
\caption{
Snapshots of model systems at equilibrium at 300 K with coverage densities of (a) ${\it\Gamma}=0.18$ and (b) ${\it\Gamma}=0.59$. 
The molecules align along the zigzag direction of graphene, maintaining an average intermolecular spacing of 4.5 \AA. 
(c) Histograms showing the distribution of angles ($\theta$) between the longitudinal orientation of the molecules and the $x$-axis during equlibration.
}
\label{fig:snapshot}
\end{figure}

The MD-predicted binding energy for a single-molecule system ($m=1$) is 1.08 eV, closely matching the value of 1.11 eV obtained from DFT calculations~\cite{Kamiya2013JJAP}, where the influence of thermal effects appears to be negligible.
As additional molecules aggregate in close proximity ($m>1$), they arrange themselves in an orderly fashion, forming stripes with an average intermolecular spacing ($\Delta l$) of approximately 4.5~\AA, as examples illustrated in Fig.~\ref{fig:snapshot}(a) and (b).
These adsorbed molecules adopt straight, all-$trans$ configurations, favoring an orientation along the zigzag direction of the graphite substrate, which aligns with experimental observations~\cite{McGonigal1990APL}.
Such a structural arrangement is energetically favorable, given that the periodicity along the graphene armchair direction is 4.26 \AA, in close agreement with $\Delta l$.

\begin{figure}[ht]
\centering
\includegraphics[width=0.5\textwidth]{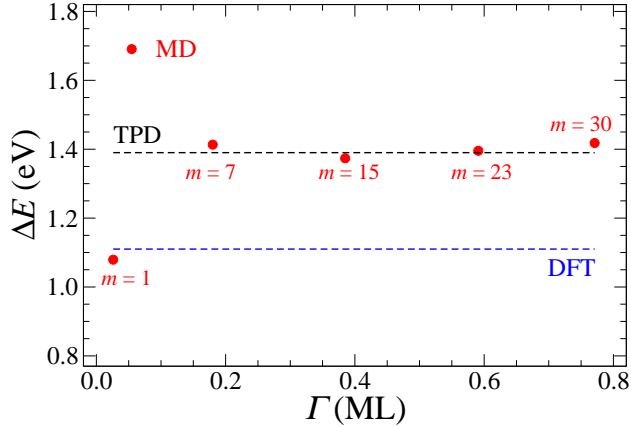}
\caption{
Binding energy ($\Delta E$) of HEX on Gr as a function of coverage density ($\it\Gamma$). 
The blue dashed line represents a DFT calculation~\cite{Kamiya2013JJAP} from the same setup. 
The value is extrapolated using the relation: $\Delta E(N_{\rm C})=-0.067N_{\rm C}-0.04$ eV, where $N_{\rm C}$ denotes the number of backbone carbons. 
The black dashed line indicates the desorption energy ($\Delta E_{\rm des}$) derived from the relationship $\Delta E_{\rm des}=-29+42N^{0.5}$, based on temperature-programmed desorption (TPD) experiments~\cite{Gellman2002JPCB}.
Error bars are omitted as they are smaller than the symbol size.
}
\label{fig:binding_energy}
\end{figure}

Histograms showing the molecular orientation during dynamic equilibrium for two representative systems are presented in Fig.~\ref{fig:snapshot}(c), where the orientation vectors are determined as the end-to-end vector of each molecule along the longitudinal direction.
An increase in binding energy of approximately 0.5 eV is observed as the number of molecules $m$ increases from one to numerous, as shown in Fig.~\ref{fig:binding_energy}.
Beyond this point, $\Delta E$ remains constant as $\it\Gamma$ further increases, regardless of the initial desorption position of the target molecule.
This suggests that the binding energies obtained here correspond to the desorption of a molecule from the domain edges, which are preferred sites due to the low desorption energy barrier caused by structural asymmetry.

The binding energies presented in Fig.~\ref{fig:binding_energy} (red solid circles) are consistent with the MD-modeled energy barriers obtained from continuous pulling of a single molecule from an adsorption state to an isolated state along the surface normal while maintaining a constant center-of-mass (COM) displacement rate of 0.05 m/s.
During this process, no reattachment of detached fragments is observed, owing to the high persistence length of these short-chain molecules.
The correlation between free energy and the number of detached fragments suggests minimal configurational isomerism from $trans$ to $gauche$ conformations.
These values are also in agreement with the desorption barrier ($\Delta E_{\rm des}$) extracted from temperature-programmed desorption (TPD) experiments (represented by the black dashed line in Fig.~\ref{fig:binding_energy}), where the linearly interpolated peak desorption temperature ($T_p$) is approximately 308 K~\cite{Paserba2001PRL, Gellman2002JPCB}, close to the 300 K used in MD simulations.
Our simulation results support the experimentally observed first-order desorption process, where the desorption barrier is independent of coverage density, indicating relatively weak and short-ranged intermolecular interactions among the molecules.
According to DFT calculations, the presence of oligomers on graphene could cause deviations in the electronic structure of graphene from the Fermi level observed in pristine graphene, resulting in an energy gap of up to 6 meV~\cite{Kamiya2013JJAP}.

\subsection{Nanoscale Confinement and Shearing}
\subsubsection{Solvation Forces}

\begin{figure}[ht]
\centering
\includegraphics[width=0.5\textwidth]{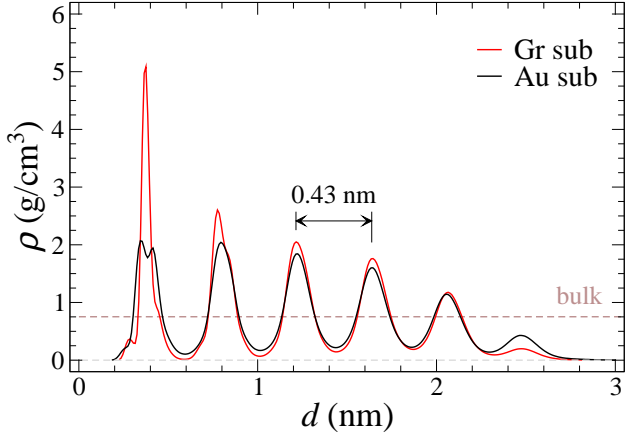}
\caption{
Density profiles ($\rho$) of HEX films on graphite (Gr, red) and gold (Au(111), black) substrates along the surface normal.
The simulation cell has an in-plane dimension of 11.5$\times$11.6 nm$^2$. 
The surface of the substrate is set at $d=0$ nm for both materials.
The HEX films exhibit a characteristic density oscillation periodicity ($\lambda$) of approximately 0.43 nm in both cases.
The number of molecules in the first monolayer is $m=117$ on Au and 128 on Gr.
}
\label{fig:density_profile}
\end{figure}

When liquids are confined within nanoscale slits, they often form layered structures~\cite{Horn1981JCP, Krass2016JPCM} as molecules self-organize with significant in-plane order due to the perturbative effect of solid surfaces.
A solid surface can induce in-plane molecular ordering that extends through several molecular layers, manifesting in oscillatory density profiles that decay with distance from the solid-liquid interface.
The in-plane ordering of HEX on Gr is notably more pronounced than on a Au(111) surface, as shown in Fig.~\ref{fig:density_profile}.
Specifically, the peak density of the HEX layer in immanent contact with Gr is nearly 2.5 times that on a Au(111) surface and 6.2 times that in a bulk phase.
This structured, densely packed, anisotropic arrangement of molecules across neighbouring layers can withstand substantial normal loads on the order of 100 MPa without necessitating a phase transformation from liquid to solid~\cite{Gao2020JCIS}.
Additionally, the prolonged relaxation time observed in confined liquids, which is inversely related to shear viscosity, leads to noticeable hysteresis in stress relaxation~\cite{Gao2022L}, underscoring the complex dynamics at play in nano-confined liquid systems.

\begin{figure}[ht]
\centering
\includegraphics[width=0.5\textwidth]{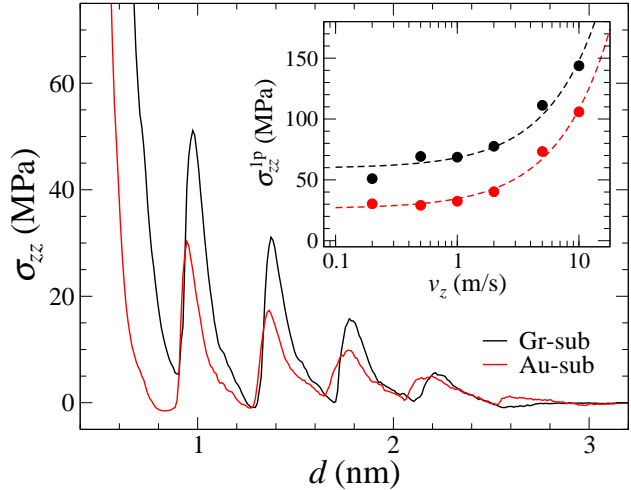}
\caption{
Normal stress ($\sigma_{zz}$) during a compression of a HEX film by a cylindrical Au(111) indenter with a radius of 3.7 nm.
The uppermost substrate surface is set at $d=0$ nm, with in-plane dimensions of 11.5$\times$11.6 nm$^2$. 
The indenter moves at a constant loading rate of $v_z=0.2$ m/s. 
The inset shows the first stress peak ($\sigma_{zz}^{\rm 1p}$) as a function of the confining rate.
}
\label{fig:sigma_zz}
\end{figure}

As a tip progressively approaches a substrate surface, the layer-by-layer squeezing out caused by in-plane molecular ordering leads to oscillatory solvation forces, with their magnitude influenced by the loading rate.
As shown in Fig.~\ref{fig:sigma_zz}, the peak normal stresses ($\sigma_{zz}^{\rm p}$) during nominally steady-state compression ($v_z=0.2$  m/s) increase exponentially as the confining distance ($d$) decreases, with a wavelength comparable to the interlayer distance.
The slightly negative stresses at the valleys correspond to the transition where molecular layers are expelled from $n$ to $n-1$.
Our results suggest that pushing the last layer of molecules (at $d<0.9$ nm) outside the contact interface to achieve solid-solid contact requires substantial effort and may even be impossible, especially when the tips are blunt, resulting in a large aspect ratio.
The heterogeneity of liquids under nanoscale confinement and the resulting stress anisotropy arise naturally from the liquid's wavelength-dependent compressibility, which aligns with bulk-liquid density autocorrelation functions (ACF)~\cite{Fisher1969JCP, Nygaard2016PRX}, rather than the specific nature of the solid surfaces.
However, detailed molecular arrangements are associated with surface topography; thus, $\sigma_{zz}$ on Gr is notably higher due to the orientation-dependent higher Gr-HEX commensurability, which induces a higher corrugation barrier, making lateral displacement more difficult.

\subsection{Shear Stress}

\begin{figure}[ht]
\centering
\includegraphics[width=0.5\textwidth]{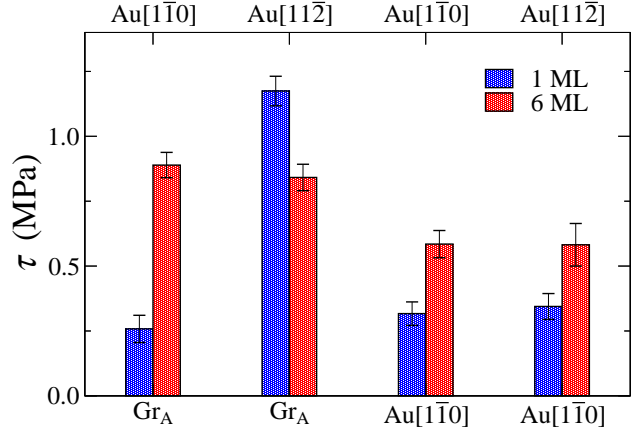}
\caption{
Shear stress ($\tau$) between two sliding slabs with an intermediate HEX film composed of either one (blue) or six (red) monolayers. 
The top and bottom ticks indicate the material types and lattice orientations aligned with the sliding direction of the respective slabs. 
The in-plane dimensions of the simulation cell are 6.4$\times$6.4 nm$^2$, ensuring that the lattice difference-induced in-plane strain is less than 1\% as an artifact.
Atoms in the outermost layers of the solids are constrained to zero stress.
The top slab slides at a constant speed of 10 m/s under only an adhesive load.
Shear stresses in the Au-HEX-Gr cases are from REF~\cite{Oo2024}.
}
\label{fig:stress}
\end{figure}

Shear stresses ($\tau$), calculated as the ratio of lateral force ($F$) to the apparent contact area ($A$), are evaluated when a Au(111) slab slides against a HEX film situated atop a Gr substrate under an adhesive load.
This loading condition is commonly employed in nanoisland manipulation experiments conducted in contact mode~\cite{Cihan2016NC, Oo2023TL, Oo2024}.
With a monolayer of HEX ($\it\Gamma\approx$ 0.8), shear stresses exhibit a strong correlation with the lattice orientation of the confining solids.
Specifically, as illustrated by the two leftmost blue bars in Fig.~\ref{fig:stress}, $\tau$ is approximately five times higher when the gold's $[11\overline{2}]$ direction vector aligns parallel to graphene's armchair direction (the sliding direction) compared to when its orthogonal vector, $[1\overline{1}0]$, aligns in the same direction.
Note that the stresses reported in Fig.~\ref{fig:stress} are from area-filling contacts, which exclude the energy dissipation at the contact lines~\cite{Gao2022FC} or due to spontaneous cluster rotation~\cite{Gao2024TL}.

\begin{figure}[ht]
\centering
\includegraphics[width=0.75\textwidth]{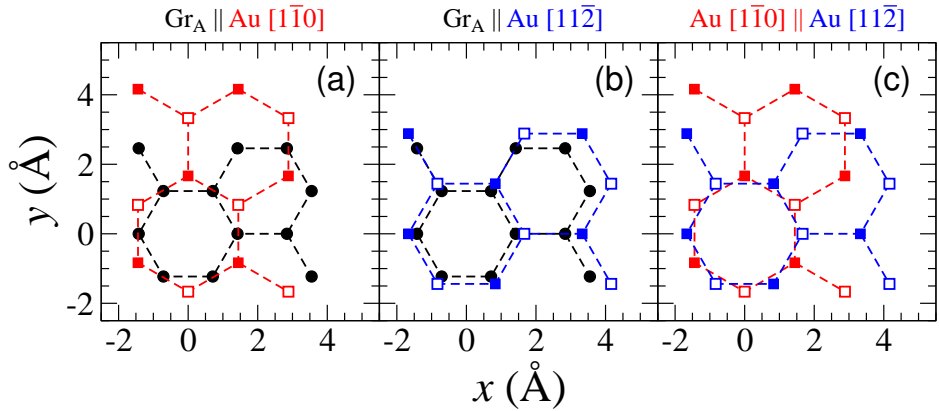}
\caption{
Illustration of lattice structures and orientations of the solid counterfaces. 
The text above each sub-panel indicates the bottom and top slabs, respectively, along with their lattice orientations, positioned on the left and right of the $\parallel$ sign. 
Solid and hollow square symbols represent Au atoms from two neighboring atomic layers. 
The sliding direction is along the $+x$ axis.
}
\label{fig:lattice}
\end{figure}

In the presence of a single monolayer of HEX molecules, the orientation-dependent shear stresses exhibit solid-like behavior, with the HEX monolayer effectively mimicking the graphene lattice.
The notable lattice congruency between Au(111) and Gr(0001), as depicted in Fig.~\ref{fig:lattice}b, facilitates the formation of Moir\'e patterns~\cite{Gao2022FC}, which, with extended superlattice wavelengths, promote longer-range instabilities at small lattice mismatches characterized by an elevated stress-strain gradient.
Although the presence of a HEX monolayer amplifies $\tau$ by nearly a hundredfold compared to gold sliding directly against graphite~\cite{Gao2022FC}, it does not cause a breakdown of superlubricity.
One reason is that this modeled large friction discrepancy would be significantly reduced at experimentally\footnote[2]{Refers specifically to nanomanipulation experiments conducted with AFM} realistic sliding speeds, which are typically nine orders of magnitude lower, due to different friction-velocity dependencies in the presence and absence of intermediate contaminants (discuss next).
Additionally, the much larger contact area of nanoislands, on the order of 10$^3\sim$ 10$^5$ nm$^2$ in experiments, will mitigate the effects of energy dissipation at contact lines caused by structural discontinuities~\cite{Gao2022FC} or molecular plowing~\cite{Flater2007L}.

%
%
%

In contrast, the orientation effect is not discernible in the Au-HEX-Au systems, as shown by the two rightmost blue bars in Fig.~\ref{fig:stress}.
This can be attributed to HEX's low binding energy to Au(111)~\cite{Wetterer1998JPCB}, which fails to foster the same degree of molecular ordering observed on graphite, even when the two solid slabs are commensurate, as evidenced by the less pronounced density peaks in Fig.~\ref{fig:density_profile}.
The orientation effect also diminishes as the HEX film thickness increases from one to six molecular layers, where viscous damping becomes increasingly dominant due to the decay of in-plane molecular ordering with distance from the solid surfaces.
As a result, the reduced interlayer lattice congruency renders the solid's lattice orientation insignificant, leading to comparable shear stresses (shown by the red bars in Fig.~\ref{fig:stress}) between systems with low and high commensurability.
However, shear stresses arising from elastic instability continue to surpass those from viscous stresses under conditions equivalent to shearing a bulk liquid in a Couette flow~\cite{Gao2024TL}.
In the latter case, although molecules align along the streaming direction according to their anisotropic radius of gyration~\cite{Gao2024TL}, their spatially-resolved density remains evenly distributed normal to the sliding plane rather than being oscillatory.

\begin{figure}[ht]
\centering
\includegraphics[width=0.46\textwidth]{stress_velocity.eps}
\includegraphics[width=0.5\textwidth]{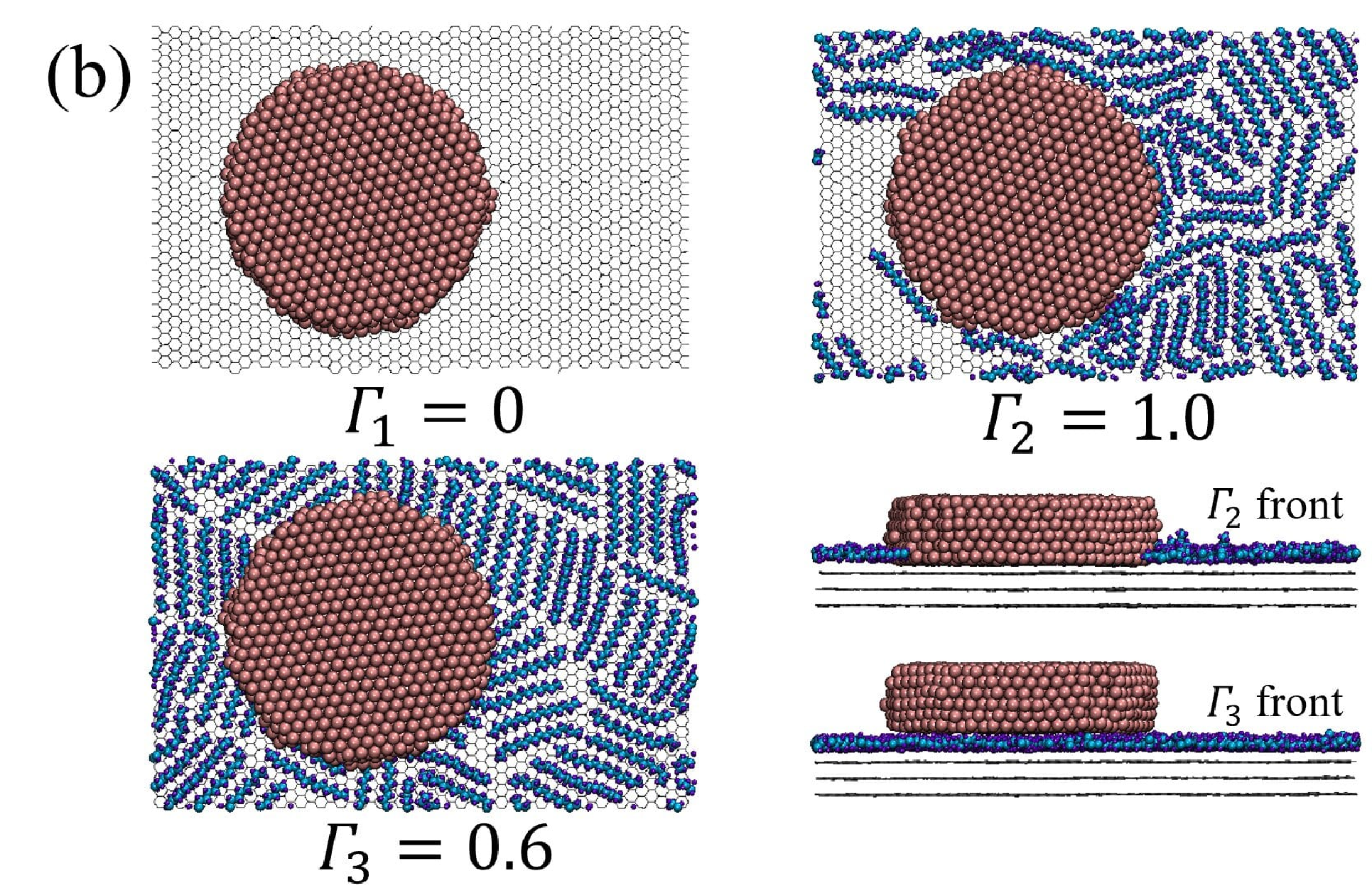}
\caption{
(a) Velocity-dependent shear stress ($\tau_{xz}$) during the sliding of a gold island on graphite (0001) at varying coverage densities ($\it\Gamma$) ranging from zero to one. 
The gold island has a cylindrical shape with a radius of 2.8 nm and a thickness of 1.0 nm.
The in-plane dimensions of the HOPG are 11.5$\times$7.4 nm$^2$.
The island slides along the armchair direction of HOPG under an adhesive load, with its (111) surface parallel to the sliding plane.
The rotational degrees of freedom of the island are allowed.
Snapshots of the model systems are shown in (b).
The dashed lines represent fitting with a power-law function, where $n$ denotes the power-law exponent.
The last data point in blue (at 50 m/s) is excluded from the fitting.
}
\label{fig:stress_velocity}
\end{figure}

With a finite geometry, contributions from both inside and outside the contact are taken into account.
Interestingly, as shown in Fig.~\ref{fig:stress_velocity}, shear stresses exhibit different dependencies on sliding speed depending on the coverage density.
In the absence of contaminants (dry contact, $\it\Gamma$ = 0), the stresses (black circles) are the lowest, with damping being Coulombic, as indicated by the power-law exponent $n$ being nearly zero.
%
%
When HOPG is lightly contaminated, meaning the island remains in direct contact with HOPG but with HEX molecules in close proximity ($\it\Gamma$ = 0.5), the shear stresses (blue triangles) are the highest, mainly due to the energy dissipated at the leading edge as HEX molecules are displaced, overcoming the accumulated corrugation barrier.
At low speeds, molecules have sufficient time to refill the sliding-induced vacancies at the trailing edge, while at high speeds, they cannot quickly flow back to the sliding trace due to hysteresis, leading to a quasi-Stokesian dependency ($n=0.67$).
In an extreme case, this can result in a drop in shear stress (at $v_x=50$ m/s) during reciprocal sliding when molecules cannot return in time at the leading edge.
When a nearly full monolayer of contaminants forms such that the island floats on it, the shear stresses (red squares) remain low and exhibit a weak dependency on sliding speed, consistent with experimental observations~\cite{Oo2024-2}.
This behavior indicates that the monolayer is in a solid-like state, with closely packed and aligned molecules, persisting superlubricity or to say quasi-structural lubricity.

\section{Conclusions}

Under ambient conditions, the adsorption of airborne molecules on graphite (Gr) is inevitable, leading to the formation of orderly arranged molecular monolayers that impact interfacial mechanical and tribological properties.
This study specifically examines $n$-hexadecane (HEX), a linear alkane, as a representative contaminant.
To enable reliable molecular dynamics predictions, the interfacial interaction between HEX and Gr was firstly parameterized using the Buckingham formalism, based on DFT-calculated molecular forces and binding energies.
This approach enhances predictions of force components, particularly in the normal direction, and thereby the corrugation energy barriers encountered during the lateral motions. 
The aim is to develop a simple yet accurate means for predicting interfacial behavior, while implicitly accounting for variations in electrostatic properties upon contact.
The robustness of the developed potential parameters was next validated by reproducing the coverage density-dependent binding energy derived from temperature-programmed desorption experiments~\cite{Gellman2002JPCB}, with errors $\leq 5\%$.
Furthermore, the study confirms that HEX molecules adopt a straight, all-$trans$ configuration aligned parallel to the graphite zigzag direction upon adsorption, consistent with experimental observations~\cite{McGonigal1990APL}.

When a HEX film with multi-layer thickness forms on Gr, the pronounced layering-like structure, evidenced by the oscillatory density peaks along the surface normal, is significantly more distinct than on Au(111).
This distinction arises from the high in-plane molecular ordering induced by the lattice similarity between the parallel-aligned HEX molecules and Gr(0001).
Such structural anisotropy can lead to a discontinuous, layer-by-layer expulsion during steady-state compression, with the last monolayer of HEX being particularly resistant to removal.
This monolayer, directly in contact with Gr, exhibits solid-like behavior despite its diffusive nature, as indicated by the orientation-dependent shear stresses observed during boundary shearing.
This behavior contrasts with systems where the commensurability is low, such as on Au(111), where shear stresses are lower and unaffected by detailed surface topography or lattice orientation, and the shear plane locates typically at such heterojunctions.

The monolayer of adsorbents can serve as a protective film for substrate surfaces, where superlubricity does not necessarily break down when accounting for the length and speed scaling from the MD scope to practical engineering applications, i.e., from $\mathcal{O}(1~{\rm nm}^2, 1\sim10~{\rm m/s})$ to $\mathcal{O}(10^2\sim10^4~{\rm nm}^2, 10^{-1}\sim10^2~{\rm \mu m/s})$.
As the coverage density increases from $\it\Gamma$ = 0 to a full monolayer, the dependencies of shear stresses on sliding velocity transition non-monotonically from Coulomb to quasi-Stokesian, and then to quasi-Coulomb, highlighting the variation in energy dissipation channels within and outside the contact.
Among these scenarios, displacing correlated molecules from the leading edge requires considerable energy, which increases nearly linearly with sliding speed up to a threshold, beyond which shear stress drops due to the lack of molecules reflow into the sliding trace during reciprocal sliding as a result of hysteresis.
Conversely, when floating on a monolayer of HEX, shear stress increases only slightly, with comparable dissipation occurring both inside and outside the contact.
As the thickness of the HEX film increases, the impact of lattice orientation diminishes due to the increasingly viscous nature of the film, shifting the lubrication scenario from the boundary lubrication (BL) to the elastohydrodynamic lubrication (EHL).

\begin{acknowledgement}

We thank Martin M\"user, Mehmet Baykara, Wengen Ouyang, and Wai Oo for useful discussions.
This research was supported by the German Research Foundation (DFG) under grant number GA 3059/2-1.

\end{acknowledgement}





\providecommand{\latin}[1]{#1}
\makeatletter
\providecommand{\doi}
  {\begingroup\let\do\@makeother\dospecials
  \catcode`\{=1 \catcode`\}=2 \doi@aux}
\providecommand{\doi@aux}[1]{\endgroup\texttt{#1}}
\makeatother
\providecommand*\mcitethebibliography{\thebibliography}
\csname @ifundefined\endcsname{endmcitethebibliography}  {\let\endmcitethebibliography\endthebibliography}{}
\begin{mcitethebibliography}{58}
\providecommand*\natexlab[1]{#1}
\providecommand*\mciteSetBstSublistMode[1]{}
\providecommand*\mciteSetBstMaxWidthForm[2]{}
\providecommand*\mciteBstWouldAddEndPuncttrue
  {\def\EndOfBibitem{\unskip.}}
\providecommand*\mciteBstWouldAddEndPunctfalse
  {\let\EndOfBibitem\relax}
\providecommand*\mciteSetBstMidEndSepPunct[3]{}
\providecommand*\mciteSetBstSublistLabelBeginEnd[3]{}
\providecommand*\EndOfBibitem{}
\mciteSetBstSublistMode{f}
\mciteSetBstMaxWidthForm{subitem}{(\alph{mcitesubitemcount})}
\mciteSetBstSublistLabelBeginEnd
  {\mcitemaxwidthsubitemform\space}
  {\relax}
  {\relax}

\bibitem[Müser(2004)]{Mueser2004EPL}
Müser,~M.~H. Structural lubricity: Role of dimension and symmetry. \emph{Europhysics Letters} \textbf{2004}, \emph{66}, 97\relax
\mciteBstWouldAddEndPuncttrue
\mciteSetBstMidEndSepPunct{\mcitedefaultmidpunct}
{\mcitedefaultendpunct}{\mcitedefaultseppunct}\relax
\EndOfBibitem
\bibitem[Adelman and Doll(1976)Adelman, and Doll]{Adelman1976JCP}
Adelman,~S.~A.; Doll,~J.~D. {Generalized Langevin equation approach for atom/solid‐surface scattering: General formulation for classical scattering off harmonic solids}. \emph{The Journal of Chemical Physics} \textbf{1976}, \emph{64}, 2375--2388\relax
\mciteBstWouldAddEndPuncttrue
\mciteSetBstMidEndSepPunct{\mcitedefaultmidpunct}
{\mcitedefaultendpunct}{\mcitedefaultseppunct}\relax
\EndOfBibitem
\bibitem[M\"user \latin{et~al.}(2001)M\"user, Wenning, and Robbins]{Mueser2001PRL}
M\"user,~M.~H.; Wenning,~L.; Robbins,~M.~O. Simple Microscopic Theory of Amontons's Laws for Static Friction. \emph{Phys. Rev. Lett.} \textbf{2001}, \emph{86}, 1295--1298\relax
\mciteBstWouldAddEndPuncttrue
\mciteSetBstMidEndSepPunct{\mcitedefaultmidpunct}
{\mcitedefaultendpunct}{\mcitedefaultseppunct}\relax
\EndOfBibitem
\bibitem[Dietzel \latin{et~al.}(2013)Dietzel, Feldmann, Schwarz, Fuchs, and Schirmeisen]{Dietzel2013PRL}
Dietzel,~D.; Feldmann,~M.; Schwarz,~U.~D.; Fuchs,~H.; Schirmeisen,~A. Scaling Laws of Structural Lubricity. \emph{Phys. Rev. Lett.} \textbf{2013}, \emph{111}, 235502\relax
\mciteBstWouldAddEndPuncttrue
\mciteSetBstMidEndSepPunct{\mcitedefaultmidpunct}
{\mcitedefaultendpunct}{\mcitedefaultseppunct}\relax
\EndOfBibitem
\bibitem[de~Wijn(2012)]{de-Wijn2012PRB}
de~Wijn,~A.~S. (In)commensurability, scaling, and multiplicity of friction in nanocrystals and application to gold nanocrystals on graphite. \emph{Phys. Rev. B} \textbf{2012}, \emph{86}, 085429\relax
\mciteBstWouldAddEndPuncttrue
\mciteSetBstMidEndSepPunct{\mcitedefaultmidpunct}
{\mcitedefaultendpunct}{\mcitedefaultseppunct}\relax
\EndOfBibitem
\bibitem[Gao and Müser(2022)Gao, and Müser]{Gao2022FC}
Gao,~H.; Müser,~M.~H. Structural lubricity of physisorbed gold clusters on graphite and its breakdown: Role of boundary conditions and contact lines. \emph{Frontiers in Chemistry} \textbf{2022}, \emph{10}\relax
\mciteBstWouldAddEndPuncttrue
\mciteSetBstMidEndSepPunct{\mcitedefaultmidpunct}
{\mcitedefaultendpunct}{\mcitedefaultseppunct}\relax
\EndOfBibitem
\bibitem[Sharp \latin{et~al.}(2016)Sharp, Pastewka, and Robbins]{Sharp2016PRB}
Sharp,~T.~A.; Pastewka,~L.; Robbins,~M.~O. Elasticity limits structural superlubricity in large contacts. \emph{Phys. Rev. B} \textbf{2016}, \emph{93}, 121402\relax
\mciteBstWouldAddEndPuncttrue
\mciteSetBstMidEndSepPunct{\mcitedefaultmidpunct}
{\mcitedefaultendpunct}{\mcitedefaultseppunct}\relax
\EndOfBibitem
\bibitem[Monti and Robbins(2020)Monti, and Robbins]{Monti2020ACSN}
Monti,~J.~M.; Robbins,~M.~O. Sliding Friction of Amorphous Asperities on Crystalline Substrates: Scaling with Contact Radius and Substrate Thickness. \emph{ACS Nano} \textbf{2020}, \emph{14}, 16997--17003, PMID: 33226231\relax
\mciteBstWouldAddEndPuncttrue
\mciteSetBstMidEndSepPunct{\mcitedefaultmidpunct}
{\mcitedefaultendpunct}{\mcitedefaultseppunct}\relax
\EndOfBibitem
\bibitem[He \latin{et~al.}(1999)He, Müser, and Robbins]{He1999S}
He,~G.; Müser,~M.~H.; Robbins,~M.~O. Adsorbed Layers and the Origin of Static Friction. \emph{Science} \textbf{1999}, \emph{284}, 1650--1652\relax
\mciteBstWouldAddEndPuncttrue
\mciteSetBstMidEndSepPunct{\mcitedefaultmidpunct}
{\mcitedefaultendpunct}{\mcitedefaultseppunct}\relax
\EndOfBibitem
\bibitem[Annett and Cross(2016)Annett, and Cross]{Annett2016N}
Annett,~J.; Cross,~G. L.~W. {Self-assembly of graphene ribbons by spontaneous self-tearing and peeling from a substrate}. \emph{Nature} \textbf{2016}, \emph{535}, 271--275\relax
\mciteBstWouldAddEndPuncttrue
\mciteSetBstMidEndSepPunct{\mcitedefaultmidpunct}
{\mcitedefaultendpunct}{\mcitedefaultseppunct}\relax
\EndOfBibitem
\bibitem[Gao and Müser(2020)Gao, and Müser]{Gao2020JCIS}
Gao,~H.; Müser,~M.~H. Why liquids can appear to solidify during squeeze-out – Even when they don’t. \emph{Journal of Colloid and Interface Science} \textbf{2020}, \emph{562}, 273--278\relax
\mciteBstWouldAddEndPuncttrue
\mciteSetBstMidEndSepPunct{\mcitedefaultmidpunct}
{\mcitedefaultendpunct}{\mcitedefaultseppunct}\relax
\EndOfBibitem
\bibitem[Gao(2022)]{Gao2022L}
Gao,~H. History-Dependent Stress Relaxation of Liquids under High-Confinement: A Molecular Dynamics Study. \emph{Lubricants} \textbf{2022}, \emph{10}\relax
\mciteBstWouldAddEndPuncttrue
\mciteSetBstMidEndSepPunct{\mcitedefaultmidpunct}
{\mcitedefaultendpunct}{\mcitedefaultseppunct}\relax
\EndOfBibitem
\bibitem[Huang \latin{et~al.}(2023)Huang, Li, Wang, Xia, Tan, Peng, Xiang, Liu, Ma, and Zheng]{Huang2023NC}
Huang,~X.; Li,~T.; Wang,~J.; Xia,~K.; Tan,~Z.; Peng,~D.; Xiang,~X.; Liu,~B.; Ma,~M.; Zheng,~Q. Robust microscale structural superlubricity between graphite and nanostructured surface. \emph{Nature Communications} \textbf{2023}, \emph{14}, 2931\relax
\mciteBstWouldAddEndPuncttrue
\mciteSetBstMidEndSepPunct{\mcitedefaultmidpunct}
{\mcitedefaultendpunct}{\mcitedefaultseppunct}\relax
\EndOfBibitem
\bibitem[Cihan \latin{et~al.}(2016)Cihan, İpek, Durgun, and Baykara]{Cihan2016NC}
Cihan,~E.; İpek,~S.; Durgun,~E.; Baykara,~M.~Z. Structural lubricity under ambient conditions. \emph{Nature Communications} \textbf{2016}, \emph{7}, 12055\relax
\mciteBstWouldAddEndPuncttrue
\mciteSetBstMidEndSepPunct{\mcitedefaultmidpunct}
{\mcitedefaultendpunct}{\mcitedefaultseppunct}\relax
\EndOfBibitem
\bibitem[Deng \latin{et~al.}(2018)Deng, Ma, Song, He, and Zheng]{Deng2018N}
Deng,~H.; Ma,~M.; Song,~Y.; He,~Q.; Zheng,~Q. Structural superlubricity in graphite flakes assembled under ambient conditions. \emph{Nanoscale} \textbf{2018}, \emph{10}, 14314--14320\relax
\mciteBstWouldAddEndPuncttrue
\mciteSetBstMidEndSepPunct{\mcitedefaultmidpunct}
{\mcitedefaultendpunct}{\mcitedefaultseppunct}\relax
\EndOfBibitem
\bibitem[Oo \latin{et~al.}(2024)Oo, Gao, Müser, and Baykara]{Oo2024}
Oo,~W.~H.; Gao,~H.; Müser,~M.~H.; Baykara,~M.~Z. Structural Lubricity and Molecular Contamination: Rejuvenation, Aging, and Friction Switches. 2024; \url{https://arxiv.org/abs/2407.03360}\relax
\mciteBstWouldAddEndPuncttrue
\mciteSetBstMidEndSepPunct{\mcitedefaultmidpunct}
{\mcitedefaultendpunct}{\mcitedefaultseppunct}\relax
\EndOfBibitem
\bibitem[Müser(2020)]{Mueser2020L}
Müser,~M.~H. Shear Thinning in the Prandtl Model and Its Relation to Generalized Newtonian Fluids. \emph{Lubricants} \textbf{2020}, \emph{8}\relax
\mciteBstWouldAddEndPuncttrue
\mciteSetBstMidEndSepPunct{\mcitedefaultmidpunct}
{\mcitedefaultendpunct}{\mcitedefaultseppunct}\relax
\EndOfBibitem
\bibitem[Gao and Müser(2024)Gao, and Müser]{Gao2024TL}
Gao,~H.; Müser,~M.~H. On the Shear-Thinning of Alkanes. \emph{Tribology Letters} \textbf{2024}, \emph{72}, 16\relax
\mciteBstWouldAddEndPuncttrue
\mciteSetBstMidEndSepPunct{\mcitedefaultmidpunct}
{\mcitedefaultendpunct}{\mcitedefaultseppunct}\relax
\EndOfBibitem
\bibitem[Zhou \latin{et~al.}(2004)Zhou, Johnson, and Wadley]{Zhou2004PRB}
Zhou,~X.~W.; Johnson,~R.~A.; Wadley,~H. N.~G. Misfit-energy-increasing dislocations in vapor-deposited CoFe/NiFe multilayers. \emph{Phys. Rev. B} \textbf{2004}, \emph{69}, 144113\relax
\mciteBstWouldAddEndPuncttrue
\mciteSetBstMidEndSepPunct{\mcitedefaultmidpunct}
{\mcitedefaultendpunct}{\mcitedefaultseppunct}\relax
\EndOfBibitem
\bibitem[Stuart \latin{et~al.}(2000)Stuart, Tutein, and Harrison]{Stuart2000JCP}
Stuart,~S.~J.; Tutein,~A.~B.; Harrison,~J.~A. {A reactive potential for hydrocarbons with intermolecular interactions}. \emph{The Journal of Chemical Physics} \textbf{2000}, \emph{112}, 6472--6486\relax
\mciteBstWouldAddEndPuncttrue
\mciteSetBstMidEndSepPunct{\mcitedefaultmidpunct}
{\mcitedefaultendpunct}{\mcitedefaultseppunct}\relax
\EndOfBibitem
\bibitem[Price \latin{et~al.}(2001)Price, Ostrovsky, and Jorgensen]{Price2001JCC}
Price,~M. L.~P.; Ostrovsky,~D.; Jorgensen,~W.~L. Gas-phase and liquid-state properties of esters, nitriles, and nitro compounds with the OPLS-AA force field. \emph{Journal of Computational Chemistry} \textbf{2001}, \emph{22}, 1340--1352\relax
\mciteBstWouldAddEndPuncttrue
\mciteSetBstMidEndSepPunct{\mcitedefaultmidpunct}
{\mcitedefaultendpunct}{\mcitedefaultseppunct}\relax
\EndOfBibitem
\bibitem[Siu \latin{et~al.}(2012)Siu, Pluhackova, and Böckmann]{Siu2012JCTC}
Siu,~S. W.~I.; Pluhackova,~K.; Böckmann,~R.~A. Optimization of the OPLS-AA Force Field for Long Hydrocarbons. \emph{Journal of Chemical Theory and Computation} \textbf{2012}, \emph{8}, 1459--1470, PMID: 26596756\relax
\mciteBstWouldAddEndPuncttrue
\mciteSetBstMidEndSepPunct{\mcitedefaultmidpunct}
{\mcitedefaultendpunct}{\mcitedefaultseppunct}\relax
\EndOfBibitem
\bibitem[de~la Rosa-Abad \latin{et~al.}(2016)de~la Rosa-Abad, Soldano, Mejía-Rosales, and Mariscal]{de-la-Rosa-Abad2016RSCA}
de~la Rosa-Abad,~J.~A.; Soldano,~G.~J.; Mejía-Rosales,~S.~J.; Mariscal,~M.~M. Immobilization of Au nanoparticles on graphite tunnels through nanocapillarity. \emph{RSC Adv.} \textbf{2016}, \emph{6}, 77195--77200\relax
\mciteBstWouldAddEndPuncttrue
\mciteSetBstMidEndSepPunct{\mcitedefaultmidpunct}
{\mcitedefaultendpunct}{\mcitedefaultseppunct}\relax
\EndOfBibitem
\bibitem[Pu \latin{et~al.}(2007)Pu, Leng, Zhao, and Cummings]{Pu2007N}
Pu,~Q.; Leng,~Y.; Zhao,~X.; Cummings,~P.~T. Molecular simulations of stretching gold nanowires in solvents. \emph{Nanotechnology} \textbf{2007}, \emph{18}, 424007\relax
\mciteBstWouldAddEndPuncttrue
\mciteSetBstMidEndSepPunct{\mcitedefaultmidpunct}
{\mcitedefaultendpunct}{\mcitedefaultseppunct}\relax
\EndOfBibitem
\bibitem[Thompson \latin{et~al.}(2022)Thompson, Aktulga, Berger, Bolintineanu, Brown, Crozier, {in 't Veld}, Kohlmeyer, Moore, Nguyen, Shan, Stevens, Tranchida, Trott, and Plimpton]{Thompson2022CPC}
Thompson,~A.~P.; Aktulga,~H.~M.; Berger,~R.; Bolintineanu,~D.~S.; Brown,~W.~M.; Crozier,~P.~S.; {in 't Veld},~P.~J.; Kohlmeyer,~A.; Moore,~S.~G.; Nguyen,~T.~D.; Shan,~R.; Stevens,~M.~J.; Tranchida,~J.; Trott,~C.; Plimpton,~S.~J. LAMMPS - a flexible simulation tool for particle-based materials modeling at the atomic, meso, and continuum scales. \emph{Computer Physics Communications} \textbf{2022}, \emph{271}, 108171\relax
\mciteBstWouldAddEndPuncttrue
\mciteSetBstMidEndSepPunct{\mcitedefaultmidpunct}
{\mcitedefaultendpunct}{\mcitedefaultseppunct}\relax
\EndOfBibitem
\bibitem[Buckingham and Lennard-Jones(1938)Buckingham, and Lennard-Jones]{Buckingham1938PRSL}
Buckingham,~R.~A.; Lennard-Jones,~J.~E. The classical equation of state of gaseous helium, neon and argon. \emph{Proceedings of the Royal Society of London. Series A. Mathematical and Physical Sciences} \textbf{1938}, \emph{168}, 264--283\relax
\mciteBstWouldAddEndPuncttrue
\mciteSetBstMidEndSepPunct{\mcitedefaultmidpunct}
{\mcitedefaultendpunct}{\mcitedefaultseppunct}\relax
\EndOfBibitem
\bibitem[Kong \latin{et~al.}(2009)Kong, Denniston, Müser, and Qi]{Kong2009PCCP}
Kong,~L.-T.; Denniston,~C.; Müser,~M.~H.; Qi,~Y. Non-bonded force field for the interaction between metals and organic molecules: a case study of olefins on aluminum. \emph{Phys. Chem. Chem. Phys.} \textbf{2009}, \emph{11}, 10195--10203\relax
\mciteBstWouldAddEndPuncttrue
\mciteSetBstMidEndSepPunct{\mcitedefaultmidpunct}
{\mcitedefaultendpunct}{\mcitedefaultseppunct}\relax
\EndOfBibitem
\bibitem[Müser(2022)]{Mueser2022MS}
Müser,~M.~H. Improved cutoff functions for short-range potentials and the Wolf summation. \emph{Molecular Simulation} \textbf{2022}, \emph{48}, 1393--1401\relax
\mciteBstWouldAddEndPuncttrue
\mciteSetBstMidEndSepPunct{\mcitedefaultmidpunct}
{\mcitedefaultendpunct}{\mcitedefaultseppunct}\relax
\EndOfBibitem
\bibitem[Lorentz(1881)]{Lorentz1881AP}
Lorentz,~H.~A. Ueber die Anwendung des Satzes vom Virial in der kinetischen Theorie der Gase. \emph{Annalen der Physik} \textbf{1881}, \emph{248}, 127--136\relax
\mciteBstWouldAddEndPuncttrue
\mciteSetBstMidEndSepPunct{\mcitedefaultmidpunct}
{\mcitedefaultendpunct}{\mcitedefaultseppunct}\relax
\EndOfBibitem
\bibitem[Berthelot(1898)]{Berthelot1898CR}
Berthelot,~D. Sur le m{\'e}lange des gaz. \emph{Compt. Rendus} \textbf{1898}, \emph{126}, 15\relax
\mciteBstWouldAddEndPuncttrue
\mciteSetBstMidEndSepPunct{\mcitedefaultmidpunct}
{\mcitedefaultendpunct}{\mcitedefaultseppunct}\relax
\EndOfBibitem
\bibitem[VandeVondele \latin{et~al.}(2005)VandeVondele, Krack, Mohamed, Parrinello, Chassaing, and Hutter]{VandeVondele2005CPC}
VandeVondele,~J.; Krack,~M.; Mohamed,~F.; Parrinello,~M.; Chassaing,~T.; Hutter,~J. Quickstep: Fast and accurate density functional calculations using a mixed Gaussian and plane waves approach. \emph{Computer Physics Communications} \textbf{2005}, \emph{167}, 103--128\relax
\mciteBstWouldAddEndPuncttrue
\mciteSetBstMidEndSepPunct{\mcitedefaultmidpunct}
{\mcitedefaultendpunct}{\mcitedefaultseppunct}\relax
\EndOfBibitem
\bibitem[{Lippert} \latin{et~al.}(1997){Lippert}, {Hutter}, and {Parrinello}]{Gerald1997MP}
{Lippert},~G.; {Hutter},~J.; {Parrinello},~M. {A hybrid Gaussian and plane wave density functional scheme}. \emph{Molecular Physics} \textbf{1997}, \emph{92}, 477--488\relax
\mciteBstWouldAddEndPuncttrue
\mciteSetBstMidEndSepPunct{\mcitedefaultmidpunct}
{\mcitedefaultendpunct}{\mcitedefaultseppunct}\relax
\EndOfBibitem
\bibitem[Hutter \latin{et~al.}(2014)Hutter, Iannuzzi, Schiffmann, and VandeVondele]{Hutter2014CMS}
Hutter,~J.; Iannuzzi,~M.; Schiffmann,~F.; VandeVondele,~J. cp2k: atomistic simulations of condensed matter systems. \emph{WIREs Computational Molecular Science} \textbf{2014}, \emph{4}, 15--25\relax
\mciteBstWouldAddEndPuncttrue
\mciteSetBstMidEndSepPunct{\mcitedefaultmidpunct}
{\mcitedefaultendpunct}{\mcitedefaultseppunct}\relax
\EndOfBibitem
\bibitem[Kühne \latin{et~al.}(2020)Kühne, Iannuzzi, Del~Ben, Rybkin, Seewald, Stein, Laino, Khaliullin, Schütt, Schiffmann, Golze, Wilhelm, Chulkov, Bani-Hashemian, Weber, Borštnik, Taillefumier, Jakobovits, Lazzaro, Pabst, Müller, Schade, Guidon, Andermatt, Holmberg, Schenter, Hehn, Bussy, Belleflamme, Tabacchi, Glöß, Lass, Bethune, Mundy, Plessl, Watkins, VandeVondele, Krack, and Hutter]{Kuhne2020JCP}
Kühne,~T.~D. \latin{et~al.}  {CP2K: An electronic structure and molecular dynamics software package - Quickstep: Efficient and accurate electronic structure calculations}. \emph{The Journal of Chemical Physics} \textbf{2020}, \emph{152}, 194103\relax
\mciteBstWouldAddEndPuncttrue
\mciteSetBstMidEndSepPunct{\mcitedefaultmidpunct}
{\mcitedefaultendpunct}{\mcitedefaultseppunct}\relax
\EndOfBibitem
\bibitem[Dion \latin{et~al.}(2004)Dion, Rydberg, Schr\"oder, Langreth, and Lundqvist]{Dion2004PRL}
Dion,~M.; Rydberg,~H.; Schr\"oder,~E.; Langreth,~D.~C.; Lundqvist,~B.~I. Van der Waals Density Functional for General Geometries. \emph{Phys. Rev. Lett.} \textbf{2004}, \emph{92}, 246401\relax
\mciteBstWouldAddEndPuncttrue
\mciteSetBstMidEndSepPunct{\mcitedefaultmidpunct}
{\mcitedefaultendpunct}{\mcitedefaultseppunct}\relax
\EndOfBibitem
\bibitem[VandeVondele and Hutter(2007)VandeVondele, and Hutter]{VandeVondele2007JCP}
VandeVondele,~J.; Hutter,~J. {Gaussian basis sets for accurate calculations on molecular systems in gas and condensed phases}. \emph{The Journal of Chemical Physics} \textbf{2007}, \emph{127}, 114105\relax
\mciteBstWouldAddEndPuncttrue
\mciteSetBstMidEndSepPunct{\mcitedefaultmidpunct}
{\mcitedefaultendpunct}{\mcitedefaultseppunct}\relax
\EndOfBibitem
\bibitem[Goedecker \latin{et~al.}(1996)Goedecker, Teter, and Hutter]{Goedecker1996PRB}
Goedecker,~S.; Teter,~M.; Hutter,~J. Separable dual-space Gaussian pseudopotentials. \emph{Phys. Rev. B} \textbf{1996}, \emph{54}, 1703--1710\relax
\mciteBstWouldAddEndPuncttrue
\mciteSetBstMidEndSepPunct{\mcitedefaultmidpunct}
{\mcitedefaultendpunct}{\mcitedefaultseppunct}\relax
\EndOfBibitem
\bibitem[Hartwigsen \latin{et~al.}(1998)Hartwigsen, Goedecker, and Hutter]{Hartwigsen1998PRB}
Hartwigsen,~C.; Goedecker,~S.; Hutter,~J. Relativistic separable dual-space Gaussian pseudopotentials from H to Rn. \emph{Phys. Rev. B} \textbf{1998}, \emph{58}, 3641--3662\relax
\mciteBstWouldAddEndPuncttrue
\mciteSetBstMidEndSepPunct{\mcitedefaultmidpunct}
{\mcitedefaultendpunct}{\mcitedefaultseppunct}\relax
\EndOfBibitem
\bibitem[Perdew \latin{et~al.}(1996)Perdew, Burke, and Ernzerhof]{Perdew1996PRL}
Perdew,~J.~P.; Burke,~K.; Ernzerhof,~M. Generalized Gradient Approximation Made Simple. \emph{Phys. Rev. Lett.} \textbf{1996}, \emph{77}, 3865--3868\relax
\mciteBstWouldAddEndPuncttrue
\mciteSetBstMidEndSepPunct{\mcitedefaultmidpunct}
{\mcitedefaultendpunct}{\mcitedefaultseppunct}\relax
\EndOfBibitem
\bibitem[Grimme \latin{et~al.}(2010)Grimme, Antony, Ehrlich, and Krieg]{Grimme2010JCP}
Grimme,~S.; Antony,~J.; Ehrlich,~S.; Krieg,~H. {A consistent and accurate ab initio parametrization of density functional dispersion correction (DFT-D) for the 94 elements H-Pu}. \emph{The Journal of Chemical Physics} \textbf{2010}, \emph{132}, 154104\relax
\mciteBstWouldAddEndPuncttrue
\mciteSetBstMidEndSepPunct{\mcitedefaultmidpunct}
{\mcitedefaultendpunct}{\mcitedefaultseppunct}\relax
\EndOfBibitem
\bibitem[Kirkpatrick \latin{et~al.}(1983)Kirkpatrick, Gelatt, and Vecchi]{Kirkpatrick1983S}
Kirkpatrick,~S.; Gelatt,~C.~D.; Vecchi,~M.~P. Optimization by Simulated Annealing. \emph{Science} \textbf{1983}, \emph{220}, 671--680\relax
\mciteBstWouldAddEndPuncttrue
\mciteSetBstMidEndSepPunct{\mcitedefaultmidpunct}
{\mcitedefaultendpunct}{\mcitedefaultseppunct}\relax
\EndOfBibitem
\bibitem[Oo \latin{et~al.}(2023)Oo, Baykara, and Gao]{Oo2023TL}
Oo,~W.; Baykara,~M.; Gao,~H. A Computational Study of Cluster Dynamics in Structural Lubricity: Role of Cluster Rotation. \emph{Tribology Letters} \textbf{2023}, \emph{71}, 115\relax
\mciteBstWouldAddEndPuncttrue
\mciteSetBstMidEndSepPunct{\mcitedefaultmidpunct}
{\mcitedefaultendpunct}{\mcitedefaultseppunct}\relax
\EndOfBibitem
\bibitem[Daw and Baskes(1984)Daw, and Baskes]{Daw1984PRB}
Daw,~M.~S.; Baskes,~M.~I. Embedded-atom method: Derivation and application to impurities, surfaces, and other defects in metals. \emph{Phys. Rev. B} \textbf{1984}, \emph{29}, 6443--6453\relax
\mciteBstWouldAddEndPuncttrue
\mciteSetBstMidEndSepPunct{\mcitedefaultmidpunct}
{\mcitedefaultendpunct}{\mcitedefaultseppunct}\relax
\EndOfBibitem
\bibitem[Steele(1973)]{Steele1973SS}
Steele,~W.~A. The physical interaction of gases with crystalline solids: I. Gas-solid energies and properties of isolated adsorbed atoms. \emph{Surface Science} \textbf{1973}, \emph{36}, 317--352\relax
\mciteBstWouldAddEndPuncttrue
\mciteSetBstMidEndSepPunct{\mcitedefaultmidpunct}
{\mcitedefaultendpunct}{\mcitedefaultseppunct}\relax
\EndOfBibitem
\bibitem[Tsuzuki and Fujii(2008)Tsuzuki, and Fujii]{Tsuzuki2008PCCP}
Tsuzuki,~S.; Fujii,~A. Nature and physical origin of CH/$\pi$ interaction: significant difference from conventional hydrogen bonds. \emph{Phys. Chem. Chem. Phys.} \textbf{2008}, \emph{10}, 2584--2594\relax
\mciteBstWouldAddEndPuncttrue
\mciteSetBstMidEndSepPunct{\mcitedefaultmidpunct}
{\mcitedefaultendpunct}{\mcitedefaultseppunct}\relax
\EndOfBibitem
\bibitem[Paserba and Gellman(2001)Paserba, and Gellman]{Paserba2001PRL}
Paserba,~K.~R.; Gellman,~A.~J. Kinetics and Energetics of Oligomer Desorption from Surfaces. \emph{Phys. Rev. Lett.} \textbf{2001}, \emph{86}, 4338--4341\relax
\mciteBstWouldAddEndPuncttrue
\mciteSetBstMidEndSepPunct{\mcitedefaultmidpunct}
{\mcitedefaultendpunct}{\mcitedefaultseppunct}\relax
\EndOfBibitem
\bibitem[Gellman and Paserba(2002)Gellman, and Paserba]{Gellman2002JPCB}
Gellman,~A.~J.; Paserba,~K.~R. Kinetics and Mechanism of Oligomer Desorption from Surfaces: n-Alkanes on Graphite. \emph{The Journal of Physical Chemistry B} \textbf{2002}, \emph{106}, 13231--13241\relax
\mciteBstWouldAddEndPuncttrue
\mciteSetBstMidEndSepPunct{\mcitedefaultmidpunct}
{\mcitedefaultendpunct}{\mcitedefaultseppunct}\relax
\EndOfBibitem
\bibitem[Londero \latin{et~al.}(2012)Londero, Karlson, Landahl, Ostrovskii, Rydberg, and Schröder]{Londero2012JPCM}
Londero,~E.; Karlson,~E.~K.; Landahl,~M.; Ostrovskii,~D.; Rydberg,~J.~D.; Schröder,~E. Desorption of n-alkanes from graphene: a van der Waals density functional study. \emph{Journal of Physics: Condensed Matter} \textbf{2012}, \emph{24}, 424212\relax
\mciteBstWouldAddEndPuncttrue
\mciteSetBstMidEndSepPunct{\mcitedefaultmidpunct}
{\mcitedefaultendpunct}{\mcitedefaultseppunct}\relax
\EndOfBibitem
\bibitem[Kamiya and Okada(2013)Kamiya, and Okada]{Kamiya2013JJAP}
Kamiya,~K.; Okada,~S. Energetics and Electronic Structures of Alkanes and Polyethylene Adsorbed on Graphene. \emph{Japanese Journal of Applied Physics} \textbf{2013}, \emph{52}, 06GD10\relax
\mciteBstWouldAddEndPuncttrue
\mciteSetBstMidEndSepPunct{\mcitedefaultmidpunct}
{\mcitedefaultendpunct}{\mcitedefaultseppunct}\relax
\EndOfBibitem
\bibitem[McGonigal \latin{et~al.}(1990)McGonigal, Bernhardt, and Thomson]{McGonigal1990APL}
McGonigal,~G.~C.; Bernhardt,~R.~H.; Thomson,~D.~J. {Imaging alkane layers at the liquid/graphite interface with the scanning tunneling microscope}. \emph{Applied Physics Letters} \textbf{1990}, \emph{57}, 28--30\relax
\mciteBstWouldAddEndPuncttrue
\mciteSetBstMidEndSepPunct{\mcitedefaultmidpunct}
{\mcitedefaultendpunct}{\mcitedefaultseppunct}\relax
\EndOfBibitem
\bibitem[Horn and Israelachvili(1981)Horn, and Israelachvili]{Horn1981JCP}
Horn,~R.~G.; Israelachvili,~J.~N. {Direct measurement of structural forces between two surfaces in a nonpolar liquid}. \emph{The Journal of Chemical Physics} \textbf{1981}, \emph{75}, 1400--1411\relax
\mciteBstWouldAddEndPuncttrue
\mciteSetBstMidEndSepPunct{\mcitedefaultmidpunct}
{\mcitedefaultendpunct}{\mcitedefaultseppunct}\relax
\EndOfBibitem
\bibitem[Krass \latin{et~al.}(2016)Krass, Gosvami, Carpick, Müser, and Bennewitz]{Krass2016JPCM}
Krass,~M.-D.; Gosvami,~N.~N.; Carpick,~R.~W.; Müser,~M.~H.; Bennewitz,~R. Dynamic shear force microscopy of viscosity in nanometer-confined hexadecane layers. \emph{Journal of Physics: Condensed Matter} \textbf{2016}, \emph{28}, 134004\relax
\mciteBstWouldAddEndPuncttrue
\mciteSetBstMidEndSepPunct{\mcitedefaultmidpunct}
{\mcitedefaultendpunct}{\mcitedefaultseppunct}\relax
\EndOfBibitem
\bibitem[Fisher and Wiodm(1969)Fisher, and Wiodm]{Fisher1969JCP}
Fisher,~M.~E.; Wiodm,~B. {Decay of Correlations in Linear Systems}. \emph{The Journal of Chemical Physics} \textbf{1969}, \emph{50}, 3756--3772\relax
\mciteBstWouldAddEndPuncttrue
\mciteSetBstMidEndSepPunct{\mcitedefaultmidpunct}
{\mcitedefaultendpunct}{\mcitedefaultseppunct}\relax
\EndOfBibitem
\bibitem[Nyg\aa{}rd \latin{et~al.}(2016)Nyg\aa{}rd, Sarman, Hyltegren, Chodankar, Perret, Buitenhuis, van~der Veen, and Kjellander]{Nygaard2016PRX}
Nyg\aa{}rd,~K.; Sarman,~S.; Hyltegren,~K.; Chodankar,~S.; Perret,~E.; Buitenhuis,~J.; van~der Veen,~J.~F.; Kjellander,~R. Density Fluctuations of Hard-Sphere Fluids in Narrow Confinement. \emph{Phys. Rev. X} \textbf{2016}, \emph{6}, 011014\relax
\mciteBstWouldAddEndPuncttrue
\mciteSetBstMidEndSepPunct{\mcitedefaultmidpunct}
{\mcitedefaultendpunct}{\mcitedefaultseppunct}\relax
\EndOfBibitem
\bibitem[Flater \latin{et~al.}(2007)Flater, Ashurst, and Carpick]{Flater2007L}
Flater,~E.~E.; Ashurst,~W.~R.; Carpick,~R.~W. Nanotribology of Octadecyltrichlorosilane Monolayers and Silicon: Self-Mated versus Unmated Interfaces and Local Packing Density Effects. \emph{Langmuir} \textbf{2007}, \emph{23}, 9242--9252\relax
\mciteBstWouldAddEndPuncttrue
\mciteSetBstMidEndSepPunct{\mcitedefaultmidpunct}
{\mcitedefaultendpunct}{\mcitedefaultseppunct}\relax
\EndOfBibitem
\bibitem[Wetterer \latin{et~al.}(1998)Wetterer, Lavrich, Cummings, Bernasek, and Scoles]{Wetterer1998JPCB}
Wetterer,~S.~M.; Lavrich,~D.~J.; Cummings,~T.; Bernasek,~S.~L.; Scoles,~G. Energetics and Kinetics of the Physisorption of Hydrocarbons on Au(111). \emph{The Journal of Physical Chemistry B} \textbf{1998}, \emph{102}, 9266--9275\relax
\mciteBstWouldAddEndPuncttrue
\mciteSetBstMidEndSepPunct{\mcitedefaultmidpunct}
{\mcitedefaultendpunct}{\mcitedefaultseppunct}\relax
\EndOfBibitem
\bibitem[Oo \latin{et~al.}(2024)Oo, Ashby, and Baykara]{Oo2024-2}
Oo,~W.~H.; Ashby,~P.~D.; Baykara,~M.~Z. Structural Superlubricity at High Sliding Speeds under Ambient Conditions. 2024; \url{https://arxiv.org/abs/2407.06971}\relax
\mciteBstWouldAddEndPuncttrue
\mciteSetBstMidEndSepPunct{\mcitedefaultmidpunct}
{\mcitedefaultendpunct}{\mcitedefaultseppunct}\relax
\EndOfBibitem
\end{mcitethebibliography}

\end{document}